\documentclass[journal]{IEEEtran}

\ifCLASSINFOpdf
\else
   \usepackage[dvips]{graphicx}
\fi\usepackage{multirow}
\usepackage{url}

\usepackage[justification=centering]{caption}

\usepackage{graphicx}
\usepackage{amsmath}
\usepackage[numbers,sort&compress]{natbib}

\usepackage{multirow}

\usepackage{scalerel}
\usepackage{tikz}
\usetikzlibrary{svg.path}

\definecolor{orcidlogocol}{HTML}{A6CE39}
\tikzset{
  orcidlogo/.pic={
    \fill[orcidlogocol] svg{M256,128c0,70.7-57.3,128-128,128C57.3,256,0,198.7,0,128C0,57.3,57.3,0,128,0C198.7,0,256,57.3,256,128z};
    \fill[white] svg{M86.3,186.2H70.9V79.1h15.4v48.4V186.2z}
                 svg{M108.9,79.1h41.6c39.6,0,57,28.3,57,53.6c0,27.5-21.5,53.6-56.8,53.6h-41.8V79.1z M124.3,172.4h24.5c34.9,0,42.9-26.5,42.9-39.7c0-21.5-13.7-39.7-43.7-39.7h-23.7V172.4z}
                 svg{M88.7,56.8c0,5.5-4.5,10.1-10.1,10.1c-5.6,0-10.1-4.6-10.1-10.1c0-5.6,4.5-10.1,10.1-10.1C84.2,46.7,88.7,51.3,88.7,56.8z};
  }
}

\newcommand\orcidicon[1]{\href{https://orcid.org/#1}{\mbox{\scalerel*{
\begin{tikzpicture}[yscale=-1,transform shape]
\pic{orcidlogo};
\end{tikzpicture}
}{|}}}}

\usepackage{hyperref}
\begin{document}


\title{Fast Steganalysis Method for VoIP Streams}
\author{Hao Yang,
ZhongLiang Yang, 
YongJian Bao, 
and YongFeng Huang, \IEEEmembership{Senior Member, IEEE}

\thanks{This paragraph of the first footnote will contain the date on which you submitted your paper for review. It will also contain support information, including sponsor and financial support acknowledgment. For example, ``This work was supported in part by the U.S. Department of Commerce under Grant BS123456.'' }
\thanks{}
\thanks{
The authors are with the Department of Electronic Engineering, Tsinghua University, Beijing 100084, China (e-mail:yanghao17@mails.tsinghua.edu.cn;).
}}


\markboth{Journal of \LaTeX\ Class Files, Vol. 14, No. 8, August 2015}
{Shell \MakeLowercase{\textit{et al.}}: Bare Demo of IEEEtran.cls for IEEE Journals}
\maketitle

\begin{abstract}
In this letter, we present a novel and extremely fast steganalysis method of Voice over IP (VoIP) streams, driven by the need for a quick and accurate detection of possible steganography in VoIP streams. We firstly analyzed the correlations in carriers. To better exploit the correlation in code-words, we mapped vector quantization code-words into a semantic space. In order to achieve high detection efficiency, only one hidden layer is utilized to extract the correlations between these code-words. Finally, based on the extracted correlation features, we used the softmax classifier to categorize the input stream carriers. To boost the performance of this proposed model, we incorporate a simple knowledge distillation framework into the training process. Experimental results show that the proposed method achieves  state-of-the-art performance both in detection accuracy and efficiency. In particular, the processing time of this method on average is only about 0.05\%  when sample length is as short as 0.1s, attaching strong practical value to online serving of steganography monitor.
\end{abstract}

\begin{IEEEkeywords}
Speech steganography, speech steganalysis, code-word correlation.
\end{IEEEkeywords}

\IEEEpeerreviewmaketitle

\section{Introduction}
\IEEEPARstart{S}{teganography} is an efficient way to hide secret messages
into seemingly innocent carriers without perceptible distortions, and hence it is popularly employed
to achieve covert communications. On the contrary, steganalysis aims to distinguish stego objects (objects containing
a secret message) from cover objects.
VoIP is a real-time service which enables users to make phone calls anywhere through IP data networks. With the popularity of instant messaging tools such as Wechat, Skype and Snapchat,  the network traffic of VoIP increased sharply. It has been widely reported that VoIP was an excellent scheme for
covert communication since it possesses many particular characteristic, such as instantaneity, a large mass of carrier data, high covert bandwidth and flexible conversation length \cite{VOIP-1,VoIP-2,31-Xiao2008An,Huang2012Steganography}. Despite being studied, practical online steganalysis tools for VoIP streams are still rare because most proposed method can not balance the detection accuracy and efficiency. Thus, it is crucial to develop a fast and efficient steganalysis tool of VoIP stream for online system.

Steganography in VoIP streams can be carried out in network protocol and payload field \cite{Huang2012Steganography,Zander2007Covert,Chen2002Quantization}. Compared with embedding secret data into network protocol, the latter one can achieve higher concealment \cite{Huang2012Steganography}. In order to reduce bandwidth consumption, VoIP often integrated low-bit-rate compressed speech coding standards such as G.729 and G.723.1. Quantization Index Modulation (QIM) \cite{Chen2002Quantization} make it possible to integrate secret data into low-bit-rate speech \cite{Hui2014Improving}. QIM based steganography incorporates information hiding mainly by introducing  QIM  algorithm to segment and encode the codebooks in the process of speech quantization which achieves high concealment and is hard to detect \cite{31-Xiao2008An,Hui2014Improving}.

Steganalysis of digital audio always follows the same pattern: directly extracting statistical features from the carrier and then conducting classification. Mel Frequency Cepstrum Coefficient (MFCC), Markov transition features and other combined features like high order moment of carriers are the main focus \cite{Kraetzer2007Mel,Dittmann2005Steganography,27-Li2017Steganalysis}. The classification module adopted mostly was support Vector Machine (SVM) classifier. Speech steganalysis follows the same intuition but pays more attention to correlation features between code-words and frames \cite{Lin2018RNN}. Researchers always extract handcraft features based on prior knowledge. Although this is simple and fast, these method performs poorly in detection accuracy. Recently, many architectures of neural networks such as Recurrent Neural Networks (RNNs) and Convolutional Neural Networks (CNNs) have been further studied to model a different level of correlations and they achieved the best results most recently \cite{Lin2018RNN, yang-cnnlstm,24-Chen2017Cnn}. However, most of these deep learning based models have high requirement both in storage resource and in computation capacity, making it hard for real-time serving,  which poses a great threat to the security of cyberspace. Thus, in this letter, we propose a light weight and high efficient steganalysis scheme to tackle this problem.

\section{Problem Definition}
Information-theoretic model for steganography was proposed by Cachin in \cite{Cachin2004An}. Given a message $E$, one can “embed” into the given cover carrier $C$  which conforms to distribution ${P_C}$ with the embedding function $\psi$  and then produces modified carrier $S$  which conforms to distribution ${P_S}$. The goal of steganography is to reduce the differences in statistical distribution of carriers before and after steganography as much as possible, which can be expressed as:
\begin{equation} d_f(P_{\mathcal {C}},P_{\mathcal {S}}) \leq \varepsilon .
\label{e1}
\end{equation}
On the contrary, steganalysis should make full use of the differences between cover and stego carrier and judge whether there were extra messages embeded in the  given carrier. Given a sample $\xi$, steganalysis can be taken as a map $\phi (\xi ):{R^d} \to \{ 0,1\}$. where $\phi(\xi) = 0$ means that $\xi$ is detected as cover and otherwise as stego. 

In VoIP stream, speech frames were first converted to Line Spectrum Frequency (LSF) coefficients. And the LSFs are encoded by Vector Quantization (VQ). Take a low bit rate codec G.729 as an example, the quantized LSF coefficients in each frame are described by a quantization codeword set $C = c1, c2, c3$ using codebooks $C1$, $C2$ and $C3$, respectively. Steganography scheme  such as QIM will have impact on the statistical distribution of these code-words. Thus, in this paper, we adopt the quantized LSF codeword as a clue for steganalysis of VoIP stream. In the application scenario, we can utilize sliding detection window \cite{Huang2011Detection} to collect one or several continuous packets each time to construct these quantized LSF coefficients sequences for steganalysis. Assume that the sample window size is $T$,  quantized LSF codewords can be expressed as ${S} = [s_1,s_2,...,s_T]$, where $s_i$ represents $i$-th speech frame. Our goal is to construct an end-to-end model $\phi (S)$ to predict a label ${\overset{\sim}y}$ for a given carrier. In the experiment, we can get many labeled samples from dataset such as $\{ ({S_1},{y_1}),({S_2},{y_2}), \ldots ,({S_N},{y_N})\}$, where $N$ is the total sample number in dataset and ${y_i}$ is the true label of an sample.

\section{Proposed Method}
\begin{figure}[!tp]
\centering
\includegraphics[width=\linewidth]{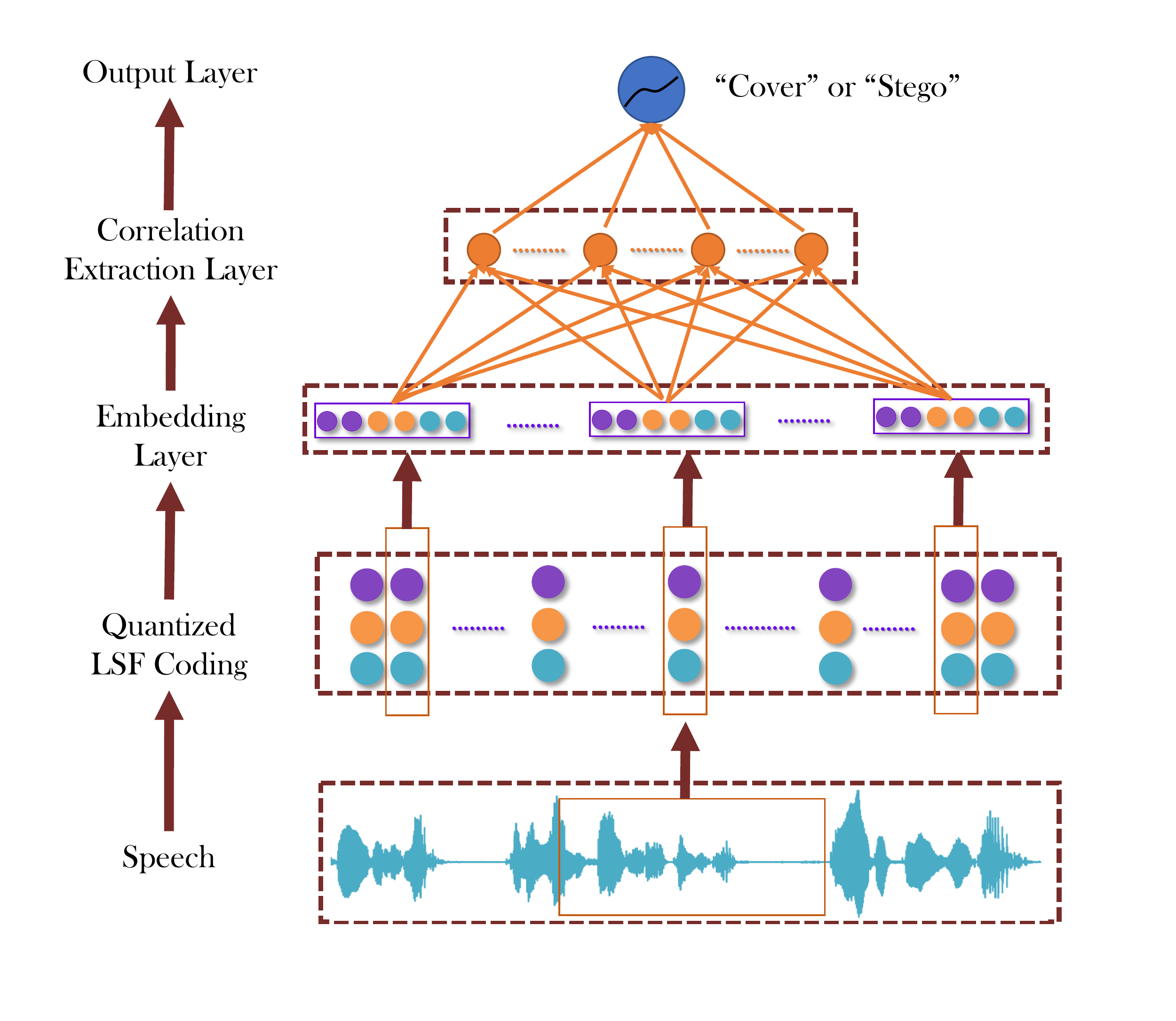}
\caption{Proposed Light Weight Correlation Extraction Scheme}
\label{fig:e}
\end{figure}

\subsection{Correlation Analysis}
There are corresponding relationships between text and speech \cite{Bosch2003Emotions}. Given a speech sequence $X$ as input, we can get an estimate of the corresponding text sequence $W$ in the following way:
\begin{equation}
\label{e1}
\begin{aligned}
{W^*} & = \mathop {{\rm{argmax}}}\limits_W P(W|X) = \mathop {{\rm{argmax}}}\limits_W {{P(X|W)P(W)} \over {P(X)}} \\
&  = \mathop {{\rm{argmax}}}\limits_W P(X|W)P(W),\qquad \quad 
\end{aligned}
\end{equation}
\noindent where  $P(X|W)$ is taken as an acoustic model  and $P(W)$ as a language model. Because of the dependency in speech sequence, word sequence and corresponding relationship between acoustic model and language model exists in Equation \ref{e1}, code-words and frames in compressed speech have strong correlation.

Define ${s_{i,j}}$ as the $i$-th codeword at frame $j$ in sequence $S$, for G.729 and G.723, $i\in [1, 3]$. When all codewords are uncorrelated, their appearances are independent. Therefore, we have:
\begin{equation} 
\label{e2}
\begin{aligned}
P({s_{i,j}} = u{\rm{ }}  & \ and \ {\rm{ }}{s_{k,l}} = v) =  P({s_{i,j}} = u)*P({s_{k,l}} = v){\rm{ }} \\
& \forall i,k \in [1,3],j,l \in [1,T],u \in {C_i},v \in {C_k}. 
\end{aligned} \end{equation}
When there are correlations between code-words ${s_{i,j}}$ and ${s_{k,l}}$, the value of two sides in  Equation \ref{e2} become different. The difference between two sides of the Equation 
\ref{e2} indicates degree of correlation. Specificly, larger imbalance of the two sides indicates stronger correlation.
In addition to explicit correlation in code-word, implicit correlation such as word correlation also exist but is difficult to model. One possible way is to map the code-words to a continuous semantic space. Recently, many remarkable results have proved that neural networks have the ability to map things like words into a continuous semantic space by self-learning with plenty of data \cite{devlin2018bert,Joulin2016Bag}. Thus, we utilized this mechanism for further study and explored semantic correlations in different levels by analyzing the distribution of these code-word vectors.

\subsection{Correlation Extraction and Classification}
Input of correlation extraction module is quantized LSF coefficients sequence $S$. At the start of correlation extraction process, we firstly map the vector quantilization code-word with a dense vector, which contains more abundant information than its original code-word. The mapping process is based on an embedding matrix $E \in {R^{{\rm{N*d}}}}$, which can be expressed as follows:
\begin{equation} 
E = \left[ \begin{array}{l}
{e_1}\\
{e_2}\\
{\rm{ }} \vdots \\
{e_N}
\end{array} \right] = \left( {\begin{array}{*{20}{c}}
{{a_{1,1}}}& \ldots &{{a_{1,d}}}\\
 \vdots & \ddots & \vdots \\
{{a_{N,1}}}& \cdots &{{a_{N,d}}}
\end{array}} \right),
\end{equation} 
where $N$ is the size of quantization code-books $C$, $d$ is the embedding dimension and the $i$-th row indicates the $i$-th code-word in $C$. Thus, we can convert vector quantilization codewords to a one-hot vector ${{\hat s}_{i,j}} \in {R^{1*N}}$ based on the value of the quantization codeword. Then, we can directly get a dense vector as follows:
\begin{equation} 
{e_{i,j}} = {{\hat s}_{i,j}}E.
\end{equation}
After the transformation, the representation of the $i$-th frame can be denoted as:
\begin{equation}
{x_i} = [{e_{i,1}} \oplus {e_{i,2}} \oplus {e_{i,3}}],
\end{equation}
where $\oplus$ is the concatenation operator. The sequence $x = {[x_1,x_2, ..., x_T]}$ was flatted to one dimension vector $h \in {R^{\rm K}}$ for further correlation analysis, where ${\rm K = 3 \times {\rm{d}} \times {\rm{T}}}$.

With these dense representations, we construct our correlation extraction 
process based on a hidden layer. Specificly, we used a weight matrix ${W_P} \in 
{R^{2 \times K}}$ to calculate the probability that this speech carrier 
contains covert information:
\begin{equation} {\left[\begin{array}{c}z_0\\ z_1 \end{array}\right]} = 
{\left[\begin{array}{c} \sum ^{K}_{t=1}w^p_{t,0}\cdot h_{t} + b_{t,0}\\ \sum 
^{K}_{t=1}w^p_{t,1}\cdot h_{t} + b_{t,1} \end{array}\right]}, 
\end{equation}
where $W_P$ and $b$ are respectively learned weight matrix and bias. The dimension of the 
output vector $z$ is 2. Finally, we add a softmax classifier to the output 
layer to calculate the possible probability of each category:
\begin{equation} T = \frac{\exp (z_{1})}{\exp (z_{0})+\exp (z_{1})}. 
\end{equation}
The output value reflects the probability that our model believes that the input sequence contains confidential messages. We can set a detection threshold, like 0.5, and then the final detection result can be expressed as
\begin{equation} {\hat y} = \left\lbrace \begin{aligned} & 1, & \left(T \geq threshold \right) \\ & 0, & \left(T < threshold \right) \end{aligned} \right. \end{equation}

\subsection{Trainning Framework}
The trainning process of the proposed model followed a supervised framework and incorporated knowledge distillation (kd) \cite{hinton2015distilling} to improve the performance. Knowledge distillation was a framework proposed by Hinton to compress a large model into a simplified model that can improve the performance of the latter one. The framework uses a teacher-student setting where the student learns from both the ground-truth labels (hard labels) and the soft labels provided by the teacher.  The probability mass associated
with each class in the soft-labels allows the student
to learn more information about the label similarities form a
given sample. 

\begin{figure}[!tp]
\centering
\includegraphics[width=\linewidth]{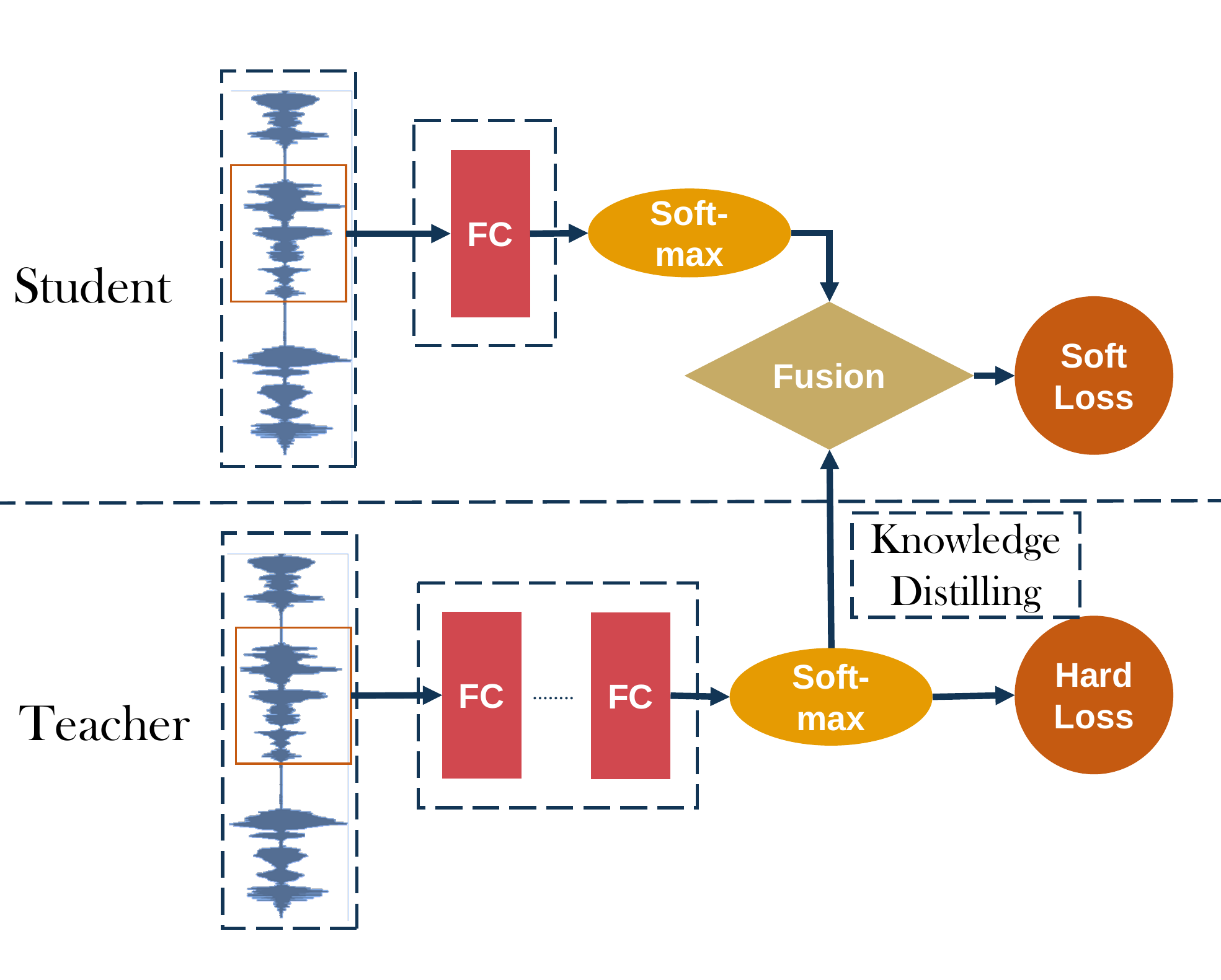}
\caption{Proposed Trainning Framework to Improve the Performance of Light Weight Correlation Extraction Scheme}
\label{fig:kd}
\end{figure}

In our setting, in order to accelerate the detection efficiency, the proposed model is designed as simple as possible. Thus, it is essential to utilized this framework in the training process of the proposed model. In this paper, we designed a teacher model which contains three Fully Connected (FC) layers to capture correlation in the sequence. Parameters in this teacher model are learned by minimizing the log loss for all training samples in the following manner:
\begin{equation}
{L^{{\rm{hard}}}} =  - \frac{1}{N}\sum\limits_{i = 1}^N {[y_i\log (\hat y_i) + (1 - y_i)\log (1 - {\hat y_i})]}, 
\label{soft}
\end{equation}
where $N$ is the total number of samples. Our key idea is to train the proposed (student) model with the resulting distribution in Equation \ref{soft} rather than the ground truth labels. Denote the label generated from teacher model as $y^{hard}$, the new loss
function for the student model can be set as:
\begin{equation}
{L^{{\rm{soft}}}} =  - \frac{1}{N}\sum\limits_{i = 1}^N {[{}y^{hard}_i}\log (\hat y_i) + (1 - {y^{hard}_i)\log (1 - {\hat y_i})]}.
\end{equation}
The framework of this training process is presented in Figure \ref{fig:kd}.

\section{Experiments and analysis}
Our experiments are based on a dataset\footnote{https://github.com/fjxmlzn/RNN-SM} that has been published by Lin $et\ al.$ \cite{Lin2018RNN}. This dataset has more than 100 hours of speech data and includes speaks both in Chinese and in English. Since different languages have different characteristics, we test the speech in different languages separately. In the experiment, we used a low-bit-rate speech codec G.729 to compress and encode the original audio. Steganography algorithm we adopted was a representative steganography scheme CNV-QIM \cite{31-Xiao2008An}. In addition, in order to test the detection performance, we cut the speech into clips with different lengths. In addition, we generated speech samples embedded with different embedding rates to test the models' adaptation to various embedding rates. Embedding rate is defined as the ratio of the number of embedded bits to the whole embedding capacity. Embedding positions are chosen randomly.

In the training process, we divided the entire sample set into training set,  validation set, and test set according to a ratio of 8:1:1. In the experiment, model hyperparameters were selected via cross-validation. In particular, the embedding size $d$ is 64. When the sample length is 0.1s, total number of frames $T$ in sliding window is 10. the number of hidden layer in proposed model is 1. The number of hidden layers in teacher model is 3 and the dimensions in last two layers are 128 and 64. We used adam \cite{adam-ref} as our learning algorithm to optimize the model parameters. Our code is implemented by Keras\footnote{https://keras.io/}. The training and testing environment for the experiments is: Inter(R) Xeon(R) CPU E5-2683 v3 \@ 2.00 GHz and GeForce GTX 1080 GPU for accelerating.

To validate the performance of our model, we chose several different representative steganalysis algorithms as our baseline models \cite{27-Li2017Steganalysis,yang-cnnlstm,Lin2018RNN}. Li $et\ al.$ \cite{27-Li2017Steganalysis} constructed a model called the Quantization codeword correlation network (QCCN) based on split VQ codeword from adjacent speech frames to capture correlation and  build a high-performance detector with the support vector machine (SVM) classifier. Lin $et\ al.$ \cite{Lin2018RNN} pointed out four types of correlation between code-words and frames and proposed a codeword correlation model, which is based on the recurrent neural network (RNN). Authors in \cite{yang-cnnlstm} indicated a proper way to take advantage of two main deep learning architectures CNN and RNN and propose a novel CNN-LSTM model to detect steganography in VoIP streams, which achieves the state-of-the-art performance. 

Besides, metrics used to evaluate the performance of different models in experiments are detection accuracy and inference time. Detection accuracy is defined as the ratio of the number of samples that are correctly classified to the total number of samples.

\subsection{Time Efficiency}
 We mainly compare our proposed method with several previous models that automatically extract correlations. The results of the experiment are shown in the Table \ref{tab:t1} and Figure \ref{fig:trainning}. From the results, it is obvious that as the length of the sample becomes longer, the detection time will continue to increase, but the detection time of our algorithm rises slowest. Moreover, our algorithm's detection performances at different sample lengths are significantly higher than other methods. When the sample length is 1s, the detection efficiency of our model is 20 times more than that of the other two models. Even if the sample length is 0.1s, inference time of proposed model is about 0.05 milliseconds  and is only 1/10 of the previous model, which shows strong  practical value for online serving.
\begin{table}[htbp]
  \centering
  \caption{Efficiency Test of Different Models}
    \begin{tabular}{l|r|r|r|r|r}
    \hline
    \multicolumn{1}{c|}{\multirow{2}[4]{*}{\textbf{Method}}} & \multicolumn{5}{c}{\textbf{Sample Length(ms)}} \\
      & \textbf{100} & \textbf{300} & \textbf{500} & \textbf{700} & \textbf{1000} \\
    \hline
    Lin  \cite{Lin2018RNN} & 0.5844 & 1.3551 & 2.1233 & 3.1001 & 4.1647 \\
    Yang  \cite{yang-cnnlstm} & 0.5964 & 1.3495 & 2.0242 & 2.7008 & 3.7536 \\
    Ours & 0.0517 & 0.0837 & 0.1001 & 0.1225 & 0.1626 \\
    \hline
    \end{tabular}%
  \label{tab:t1}%
\end{table}%




\subsection{Detection Accuracy}
\begin{table}[htbp]
  \centering
  \caption{Detection Accuracy of 0.3s Samples under Low Embedding Rates}
    \begin{tabular}{c|l|c|c|c|c|c}
    \hline
    \multicolumn{1}{c|}{\multirow{2}[4]{*}{\textbf{Language}}} & \multicolumn{1}{c|}{\multirow{2}[4]{*}{\textbf{Method}}} & \multicolumn{5}{c}{\textbf{Embedding Rates (\%) }} \\
     &   & \textbf{10} & \textbf{20} & \textbf{30} & \textbf{40} & \textbf{50} \\
    \hline
    \multirow{5}[10]{*}{\textbf{Chinese}} & Li  \cite{27-Li2017Steganalysis} & 51.85 & 58.15 & 66.25 & 71.10 & 77.40 \\
     & Lin  \cite{Lin2018RNN} & 57.04 & 67.29 & 75.41 & 82.05 & 86.13 \\
     & Yang  \cite{yang-cnnlstm} & 58.65 & 67.89 & 75.51 & 79.16 & 85.84 \\
     & Ours (no kd) & 59.69 & 69.80 & 77.80 & 84.32 & 89.21 \\
     & Ours & 60.67 & 70.18 & 78.31 & 84.59 & 89.47 \\
  \hline
    \multirow{5}[10]{*}{\textbf{English}} & Li  \cite{27-Li2017Steganalysis} & 52.30 & 58.10 & 63.05 & 72.90 & 76.30 \\
     & Lin  \cite{Lin2018RNN} & 57.18 & 66.42 & 76.74 & 81.00 & 88.25 \\
     & Yang  \cite{yang-cnnlstm} & 58.78 & 68.06 & 76.56 & 82.70 & 87.07 \\
     & Ours (no kd) & 59.84 & 69.88 & 77.71 & 84.06 & 88.99 \\
     & Ours& 60.53 & 70.32 & 78.01 & 84.44 & 89.25 \\
    \hline
    \end{tabular}%
  \label{tab:emd}%
\end{table}%

\begin{table}[htbp]
  \centering
  \caption{Detection Accuray of 20\% Embedding Rate under Short Samples}
    \begin{tabular}{c|l|c|c|c|c|c}
    \hline
    \multicolumn{1}{c|}{\multirow{2}[4]{*}{\textbf{Language}}} & \multicolumn{1}{c|}{\multirow{2}[4]{*}{\textbf{Method}}} & \multicolumn{5}{c}{\textbf{Sample Length (s)}} \\
     &   & \textbf{0.1} & \textbf{0.3} & \textbf{0.5} & \textbf{0.7} & \textbf{1.0} \\
    \hline
    \multirow{5}[10]{*}{\textbf{Chinese}} & Li  \cite{27-Li2017Steganalysis} & 54.17 & 58.15 & 61.19 & 62.33 & 65.67 \\
     & Lin  \cite{Lin2018RNN} & 58.03 & 67.29 & 72.01 & 75.08 & 77.49 \\
     & Yang  \cite{yang-cnnlstm} & 58.46 & 67.89 & 73.81 & 74.59 & 77.36 \\
     & Ours (no kd) & 61.99 & 69.80 & 74.45 & 78.12 & 81.64 \\
     & Ours & 62.27 & 70.18 & 74.76 & 78.44 & 82.06 \\
    \hline
    \multirow{5}[10]{*}{\textbf{English}} & Li  \cite{27-Li2017Steganalysis} & 57.29 & 58.10 & 61.49 & 63.48 & 66.06 \\
     & Lin  \cite{Lin2018RNN} & 57.18 & 66.79 & 73.02 & 77.32 & 80.11 \\
     & Yang  \cite{yang-cnnlstm} & 58.67 & 68.06 & 75.06 & 76.94 & 79.01 \\
     & Ours (no kd) & 61.74 & 69.88 & 74.44 & 77.68 & 81.87 \\
     & Ours & 62.27 & 70.32 & 74.89 & 78.25 & 82.31 \\
    \hline
    \end{tabular}%
  \label{tab:samplelen}%
\end{table}%
Sample duration and embedding rate are the two most important factors affecting detection performance. In a practical application scenario, two communication parties always adopt a low embedding rate strategy and communicate in a short period of time to reduce the probability of being detected. Thus, our experiments tested the performance at low embedding rates and short durations, respectively. As can be seen from the table, our model performs well at different embedding rates and different durations. In addition, it can be seen from the experiments that this proposed model utilized knowledge distillation performs better than the model that adopt commonly used training methods. Thus, we can draw the conclusion that the training framework using knowledge distillation has important value for modeling other real-time steganalysis problem.

\section{Conclusion}
In this paper, we present a novel and extremely fast steganalysis method of VoIP streams, driven by the need for a quick and accurate detection of possible steganography in stream media. Despite simple structure in exploring the correlation in the carrier, it achieves the state-of-the-art performance both in detection accuracy and efficiency. Especially, the average processing time of this proposed method is  only  about  $0.05\%$  when the sample length is as short as 0.1s, making it have strong practical value for online serving of steganography monitor. Thus, the proposed correlation extraction scheme and training framework can serve as a guide for other researchers in exploring real time steganalysis, since the solution has the potential to address similar runtime efficiency problems in online systems.
\bibliographystyle{IEEEbib}
\bibliography{strings}

\end{document}